\documentclass[final,5p,times,twocolumn]{elsarticle}

\usepackage{booktabs}
\usepackage{mathrsfs}
\usepackage{subcaption}
\usepackage{float}
\usepackage[flushleft]{threeparttable}

\usepackage{amsmath}
\usepackage{amssymb}

\journal{Nuclear Instruments and Measurements A}

\begin{document}
\DeclareGraphicsExtensions{.eps,.png,.pdf}

\begin{frontmatter}



\title{Design of a 5 GeV Laser Plasma Accelerating Module in the Quasi-linear Regime}


\author{Xiangkun Li\corref{mycorrespondingauthor}, Alban Mosnier, Phu Anh Phi Nghiem}


\address{CEA Saclay, 91191 Gif-Sur-Yvette, France}

\begin{abstract}
Multi-GeV-class laser plasma accelerating modules are key components of laser plasma accelerators, because they can be used as a booster of an upstream plasma or conventional injector or as modular acceleration sections of a multi-staged high energy plasma linac. Such a plasma module, operating in the quasi-linear regime, has been designed for the 5 GeV laser plasma accelerator stage (LPAS) of the EuPRAXIA project. The laser pulse ($\sim$150 TW, $\sim$ 15 J) is quasi-matched into a plasma channel ($n_{\rm p} = 1.5\times 10^{17}$ cm$^{-3}$, $L\sim$ 30 cm) and the bi-Gaussian electron beam is externally injected into the wakefield. The beam emittance is preserved through the acceleration by matching the beam size to the transverse focusing fields. And a final energy spread of $<$1\% has been achieved by optimizing the beam loading effect. Several methods have been proposed to reduce the slice energy spread and are found to be effective. The simulations were conducted with the 3D PIC code Warp in the Lorentz boosted frame.
\end{abstract}


\end{frontmatter}


\section{Introduction}
\label{intro}
Plasma-based accelerators have been drawing worldwide attention for decades for their capability of sustaining an accelerating gradient from a few GV/m up to 100 GV/m \cite{tajima1979laser,leemans2009laser,esarey2009physics,leemans2014multi} and are considered as potential candidates to drive compact X-ray light sources \cite{huang2012compact} or lepton colliders \cite{schroeder2010physics} in the longer term. For those high energy demands, GeV-scale plasma accelerating modules are envisaged as building blocks of a multi-staged plasma linac.

In this paper we present the design of such a plasma module operating in the quasi-linear regime for the EuPRAXIA project \cite{walker2017horizon}, where a 30 pC electron beam of 150 MeV, externally injected from a plasma or RF injector, is accelerated to 5 GeV. The beam quality should also be excellent and meet stringent requirements - low emittance ($\varepsilon_{n,x}<1~\mu$m), low energy spread ($\sigma_E/E<1\%$) and low slice energy spread ($\sigma_{E_s}/E<0.1\%$) - as it is used to drive an X-ray FEL. In the quasi-linear regime, the laser strength ($a_0 \gtrsim 1$) is not only small enough to avoid self-trapping of background electrons and to allow a better stability, but also not too small to achieve GeV's energy gain within a reasonable plasma length. The parabolic plasma channel \cite{benedetti2012quasi} was chosen to guide the laser pulse.

In this study, the key parameters of the laser pulse and the plasma channel were inferred from analytical expressions. Based on simple formulae valid in the quasi-linear regime and in the plasma channel \cite{esarey2009physics,schroeder2010physics,lu2007generating,esarey1999nonparaxial,shadwick2009nonlinear,cros2016laser}, the energy gain of the electron beam was derived by taking the dephasing, the power depletion and the self-steepening effects into account.

It will be shown later that the beam emittance can be preserved by transversely matching the beam to the linear focusing field. Although the beam loading effect introduces a non-linear transverse field, the degradation of beam emittance is small enough at the current bunch charge.  The correlated beam energy was minimized by the compensation of the wakefield curvature with the beam loading effect \cite{katsouleas1987beamloading,tzoufras2009beam}. However, in the quasi-linear regime, the beam loading also produces a nonuniform longitudinal field in the radial direction and thus induces an uncorrelated energy spread or slice energy spread.
In order to reduce it, two methods have been proposed and studied. Throughtout this study, the simulations were carried out with the 3D PIC code Warp \cite{vay2011modeling} and by using the boosted frame technique, which reduced the computation time by a factor of several hundreds.

\section{Scaling laws in the quasi-linear regime}
\label{scalings}
The operation in the quasi-linear regime requires $k_{\rm p}^2 w_0^2/2 \geq a_0^2/\gamma_{\perp}$  to avoid bubble formation and $P_L/P_c=k_{\rm p}^2 w_0^2 a_0^2/32 \leq 1$ to avoid strong relativistic self-focusing effect \cite{schroeder2010physics}, where $k_{\rm p}=\omega_{\rm p}/c=(n_{\rm p}e^2/\epsilon_0m_ec^2)^{1/2}$ is the wavenumber of the plasma wakefield, $c$ is the velocity of light in vacuum, $n_{\rm p}$ is the plasma density, $m_e$ and $e$ are the mass and charge of the electron, respectively, $\epsilon_0$ is the electric constant, $w_0$ is the spot size, $\gamma_{\perp} = (1+a_0^2/2)^{1/2}$, $P_L$ and $P_c \rm{[TW]} \approx 0.0174(\omega_0/\omega_p )^2$ are the laser power and the critical power, respectively, and $\omega_0$ the angular frequency of the laser. For the resonant pulse length, $k_{\rm p}\sigma_{\rm L} = \sqrt{2}$, with $\sigma_{L}$ the rms length of the laser amplitude, the plasma wave can be efficiently excited and the amplitude of accelerating field is given by \cite{esarey2009physics}
\begin{equation}
\label{eq:maximum-efield}
E_{z, \rm max} = 0.76a_0^2E_0/2\gamma_{\perp},
\end{equation}
where $E_0 = m_ec^2k_{\rm p}/e$ is the nonrelativistic wave breaking field. For a matched laser in a parabolic plasma channel, the phase velocity of the wake is $\beta_{\rm ph} = 1-(1/2)(\omega_{\rm p}/\omega_0)^2(1+4/k_{\rm p}^2w_0^2)$, leading to a dephasing length of $k_{\rm p}L_{\rm dp} \simeq \pi(\omega_{\rm 0}/\omega_{\rm p})^2(1+4/k_{\rm p}^2w_0^2)^{-1}$ \cite{esarey1999nonparaxial}. The power depletion length is given by $k_{\rm p}L_{\rm pd} \simeq 17.4(\omega_{\rm 0}/\omega_{\rm p})^2/a_0^2$ \cite{shadwick2009nonlinear}. The self-steepening effect is related to the power depletion length by assuming that $E_{z, \rm max}(z)/E_{z,\rm max}(0)\simeq (1+z/L_{\rm pd})$ when $z\ll L_{\rm pd}$ \cite{shadwick2009nonlinear}, with $E_{z, {\rm max}}(0)$ given by Eq.~\eqref{eq:maximum-efield}. In the quasi-linear regime, the energy gain of the electron beam is obtained by
\begin{equation}\label{eq:energy-gain-integral}
  \Delta \gamma_{\rm b} = k_{\rm p}\int_{0}^{L_{\rm acc}} \frac{E_{z,\rm max}(z)}{E_0}\cos\big(k_{\rm p}(1-\beta_{\rm ph})z\big){\rm d}z,
\end{equation}
where $\gamma_{\rm b}$ is the Lorentz factor of the beam and $L_{\rm acc}$ is the plasma length. Integration of Eq.~\eqref{eq:energy-gain-integral} leads to
\begin{align}\label{eq:energy-gain-integrated}
  \Delta \gamma_{\rm b} =& \frac{2}{\pi} k_{\rm p} L_{\rm dp} \frac{E_{z,\rm max}(0)}{E_0} \bigg[ \Big( 1+\frac{L_{\rm acc}}{L_{\rm pd}} \Big)\sin\Big(\frac{\pi}{2}\frac{L_{\rm acc}}{L_{\rm dp}}\Big)+ \nonumber\\
   &\frac{2}{\pi} \frac{L_{\rm dp}}{L_{\rm pd}} \Big( \cos\Big(\frac{\pi}{2}\frac{L_{\rm acc}}{L_{\rm dp}}\Big)-1\Big)\bigg].
\end{align}
The maximum energy gain is achieved when the beam is accelerated over the full dephasing length ($L_{\rm acc}=L_{\rm dp}$),
\begin{equation}\label{eq:energy-gain-maximum}
  \Delta \gamma_{\rm b}^{\rm max} = \frac{2}{\pi} k_{\rm p} L_{\rm dp} \frac{E_{z,\rm max}(0)}{E_0} \bigg[ 1+\frac{L_{\rm dp}}{L_{\rm pd}} \Big( 1 - \frac{2}{\pi} \Big) \bigg].
\end{equation}

\begin{figure}[H]
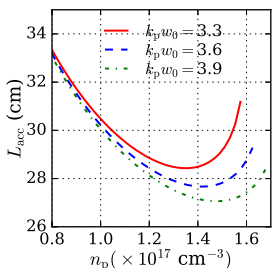

  \centering
  \begin{minipage}[b]{0.5\columnwidth}
  \centering
  \includegraphics[width = \textwidth]{{{Fig-1}}}
  \caption{Plasma length vs. its density}
  \label{fig:Lacc-np-kpw0}
  \end{minipage}
\end{figure}

Considering the EuPRAXIA project, the energy gain is 4.85 GeV. For the operation in the quasi-linear regime $a_0=\sqrt{2}$ was chosen, giving a spot size of $1.7\leq k_{\rm p}w_0\leq 4$. Then by solving Eq.~\eqref{eq:energy-gain-integrated} with $L_{\rm acc}$  as an unknown variable, we can obtain the required plasma length as a function of its density. It turns out that there is a plasma density that minimizes the plasma length, as shown in Fig.~\ref{fig:Lacc-np-kpw0}. The optimal plasma density is around $1.5\times 10^{17}$ cm$^{-3}$, corresponding to a plasma length around 30 cm. The typical parameters for the EuPRAXIA accelerating module are listed in Table~\ref{tab:typical-LPA-parameters}, where a not so large spot size was chosen to reduce the plasma size and thus the computation time.

\begin{table}[htbp]
\caption{Typical parameters for the EuPRAXIA Laser plasma acceleration stage (*: will be discussed in the following sections)}
\label{tab:typical-LPA-parameters}
\centering
\begin{tabular}{p{100pt}p{50pt}p{50pt}}
\toprule
  variable & value  & unit \\
\midrule
\textbf{Laser}   &     &                    \\
~~strength $a_0$  &  $\sqrt{2}$     & \\
~~spot size $k_{\rm p}w_0$ &  $3.3$     &          \\
~~duration $k_{\rm p}\sigma_{\rm L}$  &  $\sqrt{2}$       &      \\
~~peak power $P_{\rm L}$ & $\sim 150$ & TW \\
~~energy $E_{\rm L}$ & $\sim 15$ & J \\
\textbf{Plasma}   &      &                 \\
~~density $n_{\rm p}$ &  $1.5$ & 10$^{17}$ cm$^{-3}$ \\
~~acc. length $L_{\rm acc}$ & $\sim 30$ & cm \\
~~channel depth $\Delta n/\Delta n_c$ & $<1$* & \\
\textbf{Electron} & & \\
~~charge $Q$ & 30 & pC  \\
~~energy $E_k$ & 150 & MeV \\
~~energy spread $\Delta E/E$ & 0.5 & \% \\
~~beam size $\sigma_x$ & $\sim$1* & $\mu$m \\
~~emittance $\varepsilon_{n,x}$ & 1.0 & $\pi$ mmmrad \\
~~bunch length $\sigma_z$ & $1-3$* & $\mu$m \\
\bottomrule
\end{tabular}
\end{table}

\section{Guiding of laser pulse in a plasma channel}
\label{guiding}

The propagation of a laser pulse over a few tens of centimeters in the plasma requires the use of an optical guiding technique, for instance, a capillary tube or a preformed plasma channel. Previous studies on acceleration with capillary tubes \cite{paradkar2013numerical} have shown that the excitation of higher modes when the laser pulse is coupled into a capillary tube would result in a modulation of the transverse plasma wakefield, which in turn defocuses the electrons quickly and leads to beam losses on the capillary wall. Possible remedies have been proposed, for example the operation in the non-linear regime far from the resonance condition or the coupling of a laser pulse with a Bessel transverse distribution in order to excite only the fundamental mode. In this study, the preformed plasma channel is used to guide the laser pulse. The plasma channel has a parabolic transverse density distribution that is described by
\begin{equation}
  n_{\rm p}(r) = n_{0} \Big( 1+\frac{\Delta n}{n_0}\frac{r^2}{r_0^2} \Big)
\end{equation}
where $n_0$ is the on-axis plasma density and $\Delta n$ is called the channel depth.
A low-power ($P\ll P_c$ ) and low-intensity ($a_0 \ll 1$) laser pulse propagates without spot and intensity oscillations when the spot size and the channel depth fulfill the matching conditions, $w _0=r_0$ and $\Delta n=\Delta n_c$, where $\Delta n_c=(\pi r_e r_0^2 )^{-1}$ is called the critical channel depth and $r_e=e^2 (4\pi\epsilon_0 m_e c^2)^{-1}$ is the classical electron radius.

\begin{figure}[htbp]
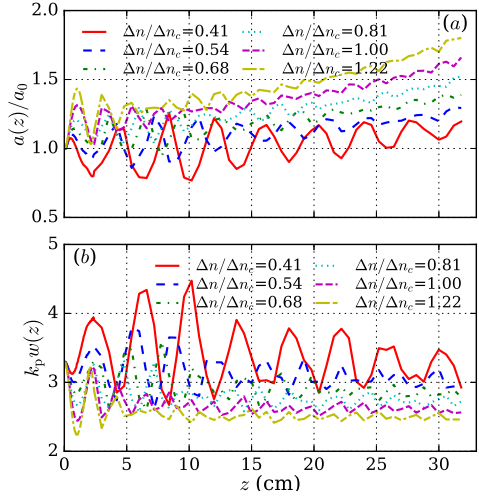

  \centering
  \begin{minipage}[b]{0.9\columnwidth}
  \centering
  \includegraphics[width = 0.85\textwidth]{{{Fig-2a}}}\\
  \includegraphics[width = 0.85\textwidth]{{{Fig-2b}}}
  \caption{Evolutions of (a) laser strength and (b) laser spot size in the plasma channel for various channel depths}
  \label{fig:laser-guiding}
  \end{minipage}
\end{figure}

In the case of a higher power ($P\lesssim P_c$) and/or intensity ($a_0\sim1$) laser, the channel depth for matched propagation should be reduced, owing to the extra-guiding coming from relativistic self-focusing and ponderomotive self-channeling \cite{benedetti2012quasi}.
A scan of the channel depth shows that the laser pulse can be quasi-matched for $\Delta n/\Delta n_c \approx 0.6$ (Fig.~\ref{fig:laser-guiding}) where both laser strength and spot size have damped oscillations around their injected values. The laser strength is slightly increased due to the self-steepening effect.

\begin{figure}[H]
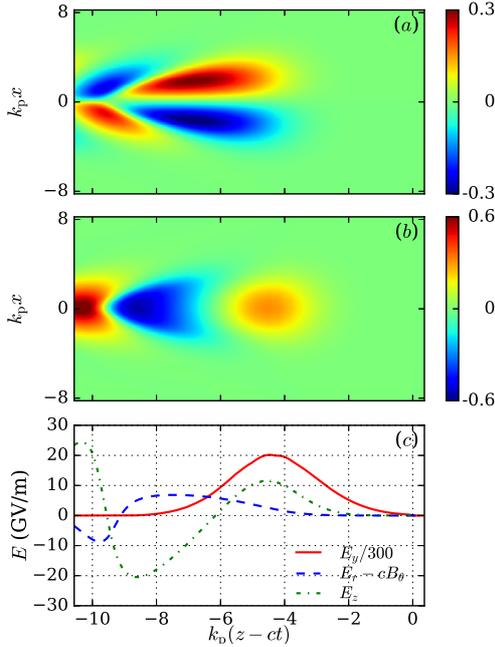

  \centering
  \begin{minipage}[b]{0.9\columnwidth}
  \centering
  \includegraphics[width = 0.85\textwidth]{{{Fig-3a}}}\\
  \includegraphics[width = 0.85\textwidth]{{{Fig-3b}}}\\
  \includegraphics[width = 0.85\textwidth]{{{Fig-3c}}}
  \caption{(a) and (b) are density plots of $E_r$ and $E_z$ respectively; (c) shows the on-axis $E_y$ (red line), $E_z$ (green dash-dotted), and $E_r-cB_{\theta}$ (blue dashed) at $r=w_0/4$}
  \label{fig:plasma-wake}
  \end{minipage}
\end{figure}

The wakefields excited by the laser in the plasma channel ($\Delta n/\Delta n_c = 0.68$) are shown in Fig.~\ref{fig:plasma-wake}, with the fields in the density plot normalized to $E_0$. The accelerating field amplitude is 20 GV/m, agreeing very well with Eq.~\eqref{eq:maximum-efield}. The laser plasma interaction is well in the quasi-linear regime, i.e. no longer in the pure linear regime but still far from the bubble regime.

\section{Preservation of beam emittance}
\label{matching}
The plasma wakefield provides a linear transverse focusing force near the beam axis, meaning that the beam emittance could be preserved if the beam size is matched. The matched size can be derived as follows \cite{reiser2008theory}
\begin{gather}\label{eq:transverse-matching}
  \sigma_x^2 = \beta_m\frac{\varepsilon_{n,x}}{\gamma_{\rm b}},~~
  \frac{1}{\beta_{m}^2} = \frac{e}{\gamma_{\rm b}m_ec^2}\frac{\partial(E_r-c B_{\theta}))}{\partial r}
\end{gather}
where $\beta_m$ is the betatron amplitude of the beam, and $E_r-cB_{\theta}$ is the transverse field that can be evaluated from simulations. Without the beam, the transverse field is linear within the range $\pm5~\mu$m and can be easily inferred from a linear fit (red points and dashed blue line in Fig.~\ref{fig:transverse-matching}). For $\varepsilon_{n,x}=1$ $\mu$m, the matched beam size is 1.3 $\mu$m. When the beam ($Q=30$ pC) is present, the beam loading (i.e. the wakefield driven by the beam itself) introduces additional nonlinear transverse field (green line in Fig.~\ref{fig:transverse-matching}) and results in an emittance growth. Figure~\ref{fig:beam-emittance-evolution} shows the final beam emittance as a function of the beam size. The lowest emittance (with a growth of only 3\%) was found for $\sigma_x = 1.4~\mu$m, very close to the theoretical one.

\begin{figure}[htbp]
  \centering
  \begin{minipage}[t]{0.5\columnwidth}
  \centering
  \includegraphics[width = 1\textwidth]{{{Fig-4}}}
  \caption{Transverse focusing force~~\\ with and w/o beam loading}
  \label{fig:transverse-matching}
  \end{minipage}%
  \begin{minipage}[t]{0.5\columnwidth}
  \includegraphics[width = 1\textwidth]{{{Fig-5}}}
  \caption{Beam emittance vs. beam size}
  \label{fig:beam-emittance-evolution}
  \end{minipage}
\end{figure}

\section{Optimization of beam energy spread}
\label{optimi}
\begin{figure}[htbp]
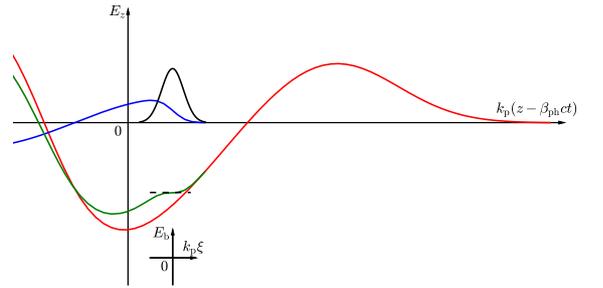

  \centering
  \begin{minipage}[b]{0.9\columnwidth}
  \centering
  \includegraphics[width = \textwidth]{{{Fig-6}}}
  \end{minipage}
  \caption{Superimposition of the fields driven by the laser (red) and by the beam (blue) produces a nearly constant field (green) within the bunch (black)}
  \label{fig:beam-loading-effect}
\end{figure}

There are several sources of energy spread. First, the correlated energy spread is induced by the wakefield curvature and the longitudinal beam loading effect, as illustrated in Fig.~\ref{fig:beam-loading-effect}. They can cancel each other out by choosing a proper bunch profile \cite{katsouleas1987beamloading,tzoufras2009beam}. 

\begin{figure}[htbp]
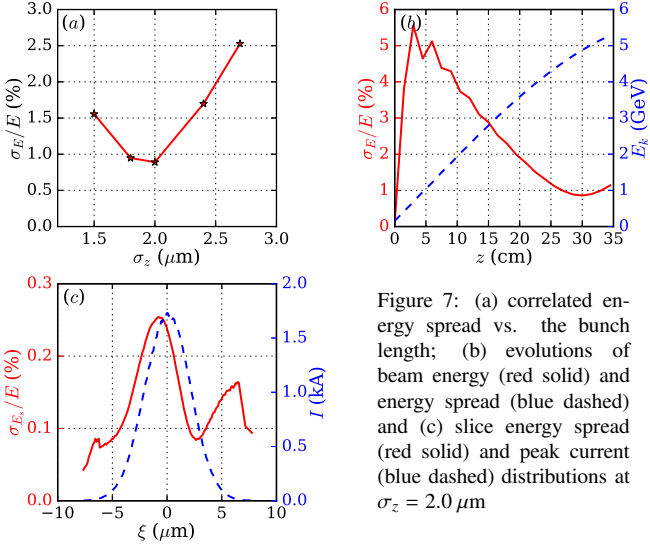

  \centering
  \begin{minipage}[b]{0.5\columnwidth}
  \centering
  \includegraphics[width = 1\textwidth]{{{Fig-7a}}}
  \end{minipage}%
  \begin{minipage}[b]{0.5\columnwidth}
  \centering
  \includegraphics[width = 1\textwidth]{{{Fig-7b}}}
  \end{minipage}
  \begin{minipage}[b]{0.5\columnwidth}
  \centering
  \includegraphics[width = 1\textwidth]{{{Fig-7c}}}
  \end{minipage}%
  \begin{minipage}[b]{0.5\columnwidth}
  \centering
  \begin{minipage}[b]{0.75\textwidth}
  \centering
  \caption{(a) correlated energy spread vs. the bunch length; (b) evolutions of beam energy (red solid) and energy spread (blue dashed) and (c) slice energy spread (red solid) and peak current (blue dashed) distributions at $\sigma_z=2.0$ $\mu$m}
  \label{fig:bunch-length-opt}
  \end{minipage}
  \end{minipage}
\end{figure}

For an initial energy spread of 0.5 \%, we scanned the bunch length to balance the wakefield and the beam loading effect. The injection phase was fixed at the crest of the wakefield. The correlated energy spread at the exit of the plasma channel (beam energy of 5 GeV) as a function of the bunch length is shown in Fig.~\ref{fig:bunch-length-opt}(a). It is found that the optimal bunch length is $\sigma_z = 2.0~\mu$m, corresponding to an energy spread lower than 1\%. The evolutions of the beam energy and energy spread along the plasma channel for this bunch length are shown in Fig.~\ref{fig:bunch-length-opt}(b).

In addition, the laser-driven wakefield as well as the longitudinal beam loading have radial dependences which will induce an uncorrelated energy spread, also called slice energy spread. In a uniform plasma, the laser-driven wakefield has the form of $E_z(z)\exp(-2r^2/w_0^2)$ \cite{esarey2009physics} and can $r$-independent near the axis when the laser spot size is much larger than the beam size. However, the radial dependence of the longitudinal beam loading can be significant and induce too high slice energy spreads for using the beam as an efficient FEL driver. Figure~\ref{fig:bunch-length-opt}(c) shows the distribution of the slice energy spread at the exit of the plasma channel, with the maximum of 0.27 \% appearing near the bunch center.

\begin{figure}[htbp]
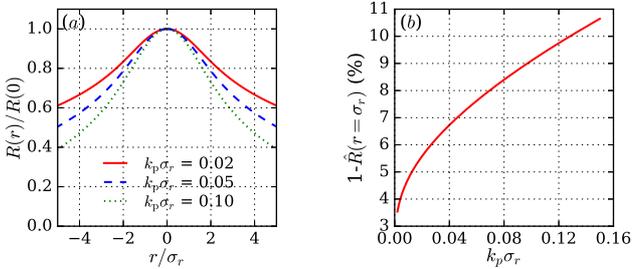

  \centering
  \begin{minipage}[b]{0.5\columnwidth}
  \centering
  \includegraphics[width = \textwidth]{{{Fig-8a}}}
  \end{minipage}%
  \begin{minipage}[b]{0.5\columnwidth}
  \centering
  \includegraphics[width = \textwidth]{{{Fig-8b}}}
  \end{minipage}
  \caption{(a) radial dependence of $E_{\rm b}$ and (b) $1-\hat{R}(r)$ as a function of $k_{\rm p}\sigma_r$}
  \label{fig:transverse-dependence}
\end{figure}

The accelerating field experienced by one particle is the sum of the laser-driven wakefield and the beam-driven wakefield (see Fig.~\ref{fig:beam-loading-effect}), $E_{\rm acc}(\xi, r) = E_{z}\big(z+\xi\big) + E_{\rm b}(\xi,r)$, where $z$ is the longitudinal coordinate of the reference particle in the laboratory frame, $\xi$ is the longitudinal coordinate within the co-moving bunch. The beam-driven wakefield is $E_{\rm b}(\xi,r)=E_{\rm b}(\xi)\hat{R}(r)$, where $\hat{R}(r) = R(r)/R(0)$ represents the radial dependence and $R(r) = \big(k_{\rm p}^2/2\pi\big)\int_{0}^{2\pi}d\theta\int_{0}^{\infty}r^{\prime}dr^{\prime}n_{\perp}(r^{\prime})K_0\big(k_{\rm p}|\vec{r}-\vec{r}^{\prime}|\big)$ from the linear theory \cite{lu2005limits}, with $n_{\perp}$ the radial distribution of the beam and $K_0$ the modified zero-order Bessel function. The function $R(r)$ for the plasma density of $1.5\times10^{17}$ cm$^{-3}$ is shown in Fig.~\ref{fig:transverse-dependence}(a) for different radial beam sizes.

The energy gain deviation between an off-axis electron and an on-axis electron is ${\rm d}E(\xi, r) = e E_{\rm b}(\xi)\big[ 1-\hat{R}(r) \big] {\rm d}z$. The final relative energy deviation is obtained by integration:
\begin{equation}\label{eq:slice-energy-diff}
  \frac{\Delta E(\xi, r)}{E} = \frac{\int_{0}^{L_{\rm acc}} e E_{\rm b}(\xi)\big[ 1-\hat{R}(r) \big] {\rm d}z}{\int_{0}^{L_{\rm acc}}e\big[E_{z}(z+\xi)+E_{\rm b}(\xi)\big]{\rm d}z} \simeq \frac{eE_{\rm b}(\xi)}{\Delta W_{\rm b}/L_{\rm acc}} \cdot \Big[ 1-\hat{R}(r) \Big],
\end{equation}
where we assume a "frozen" bunch shape during the acceleration as the incoming beam energy is sufficiently high and $\Delta W_{\rm b}$ is the energy gain of the reference particle. In Fig.~\ref{fig:transverse-dependence}(b), the quantity $1-\hat{R}(r=\sigma_r)$ is plotted as a function of the normalized beam size.
Eq.~\eqref{eq:slice-energy-diff} clearly shows that i) the slice energy spread is peaked when $E_{\rm b}(\xi)$ reaches its maximum and ii) it can be improved by reducing the normalized beam size as Fig.~\ref{fig:transverse-dependence}(b) implies and therefore by reducing the beam emittance, according to Eq.~\eqref{eq:transverse-matching}.
In addition, both $E_{\rm b}(\xi)$ and $\Delta W_{\rm b}/L_{\rm acc}$ in Eq.~\eqref{eq:slice-energy-diff} depend on the plasma density, meaning that the slice energy spread might also be reduced by tuning the plasma density.

\begin{figure}[htbp]
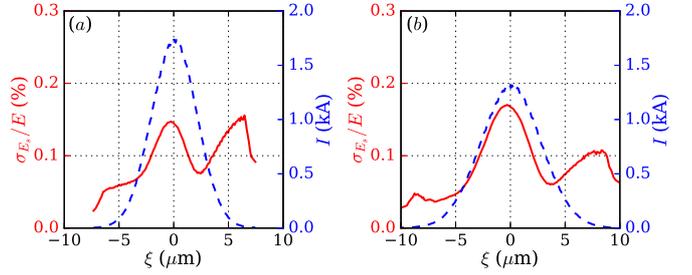

  \centering
  \begin{minipage}[b]{0.5\columnwidth}
  \centering
  \includegraphics[width = \textwidth]{{{Fig-9a}}}%
  \end{minipage}%
  \begin{minipage}[b]{0.5\columnwidth}
  \centering
  \includegraphics[width = \textwidth]{{{Fig-9b}}}
  \end{minipage}
  \caption{Slice energy spread (red solid) and peak current (blue dashed) for (a) $\varepsilon_{n,x} = 0.5~\mu$m and (b) $n_{\rm p}$ = 1.0$\times$10$^{17}$ cm$^{-3}$}\label{fig:slice-energy-spread}
\end{figure}

Simulations have been carried out with either a reduced beam emittance or a lower plasma density. The results are shown in Fig.~\ref{fig:slice-energy-spread}. When the beam emittance is reduced from 1.0 $\mu$m to 0.5 $\mu$m, the maximum slice energy spread is reduced by 38\%. When the plasma density is reduced from 1.5$\times$10$^{17}$ cm$^{-3}$ to 1.0$\times$10$^{17}$ cm$^{-3}$, the maximum slice energy spread is reduced by 33\%, with the plasma length increased from 31.5 to 35 cm. Further reduction of the plasma density sees an increase of the slice energy spread again.

\section{Conclusions}
\label{conclu}
The design and optimization of a 5 GeV laser plasma accelerator in the quasi-linear regime was presented. The laser pulse was guided in a 30 cm long parabolic plasma channel, exciting an accelerating gradient of $\sim$20 GV/m. The beam was optimized both transversely and longitudinally. By optimizing the beam size, low emittance was preserved throughout the acceleration. And by optimizing the bunch length, low energy spread ($\sigma_E/E<1$\%) was achieved. Special care was given to the slice energy spread which could be a key issue for the beam to drive X-ray FELs. Preliminary results have demonstrated that it could be improved by reducing the beam emittance or choosing a lower plasma density. While it is very difficult to reduce the beam emittance, futher studies will be carried out on the optimization of the laser-plasma parameters, for instance the plasma density and the laser strength to achieve a slice energy spread small enough for X-ray FELs.

This work has received funding from the European Union's Horizon 2020 research and innovation programme under grant agreement No. 653782.

\section*{References}
\bibliographystyle{elsarticle-num}


%
%
%


\end{document}